\documentclass[11pt,twoside]{article}

\usepackage{asp2004}
\usepackage{epsf}
\usepackage{psfig}
\usepackage{lscape}

\begin{document}
\title{The compact planetary nebula B[e] star Hen 2-90}   
\author{Michaela Kraus$^{1}$, Marcelo Borges Fernandes$^{2}$, Francisco X. de Ara\'ujo$^{3}$ and Henny J.G.L.M. Lamers$^{1}$}   
\affil{$^{1}$Sterrekundig Instituut, Utrecht University, Princetonplein 5, 3584 CC Utrecht, The Netherlands \\
$^{2}$Observat\'{o}rio do Valongo (UFRJ), Ladeira do Pedro Ant\^onio 43, 20080-090 Rio de Janeiro, Brazil\\
$^{3}$Observat\'{o}rio Nacional-MCT, Rua General Jos\'{e} Cristino 77, 20921-400 S\~{a}o Cristov\~{a}o, Rio de Janeiro, Brazil}    

\begin{abstract} 
We present a study of the optical spectrum of the fascinating B[e] star Hen 2-90
based on new high-resolution observations taken with FEROS at the ESO 1.52m 
telescope in La Silla (Chile). 
The recent HST image of Hen 2-90 \citep{Sahai02} reveals a bipolar,
highly ionized region, a neutral disk-like structure seen almost perfectly 
edge-on, and an intermediate region of moderate ionization. The slits of our 
observations cover the same innermost region of Hen 2-90 as the HST image, 
which allows us to combine the observations. Our spectra contain a huge 
amount of permitted and forbidden emission lines of atoms in different
stages of ionization. In addition, the line wings deliver velocity information
of the emitting region. We find correlation between the different ionization 
states of the elements and the velocities derived from the line profiles: the 
highly ionized atoms have the highest outflow velocity, while the neutral lines 
have the lowest. When combining the velocity information with the HST image of 
Hen 2-90, it seems that a non-spherical stellar wind model is a good option to
explain the ionization and spatial distribution of the circumstellar material.
Our modeling of the forbidden emission lines results in strong evidence
for Hen 2-90 being a compact planetary nebulae that has undergone
a superwind phase of high, non-spherical mass loss, most probably triggered
by a central star that was rotating with about 80\% of the critical 
velocity. We find a total mass loss rate during this superwind phase 
on the order of $3\times 10^{-5}$\,M$_{\odot}$yr$^{-1}$.
\end{abstract}


\begin{figure}[!th] 
\plotfiddle{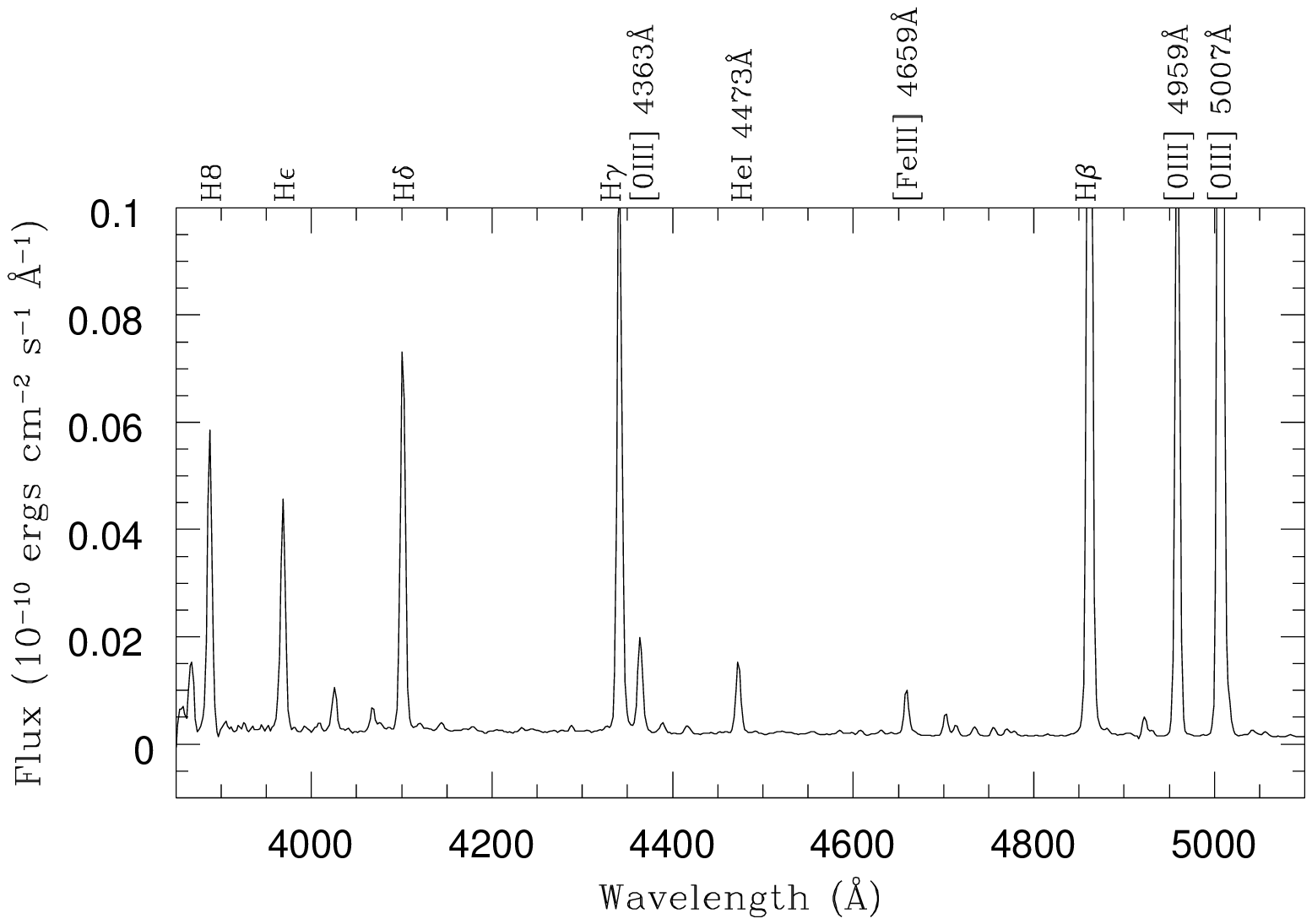}{9cm}{0}{70}{55}{-205}{0}
\plotfiddle{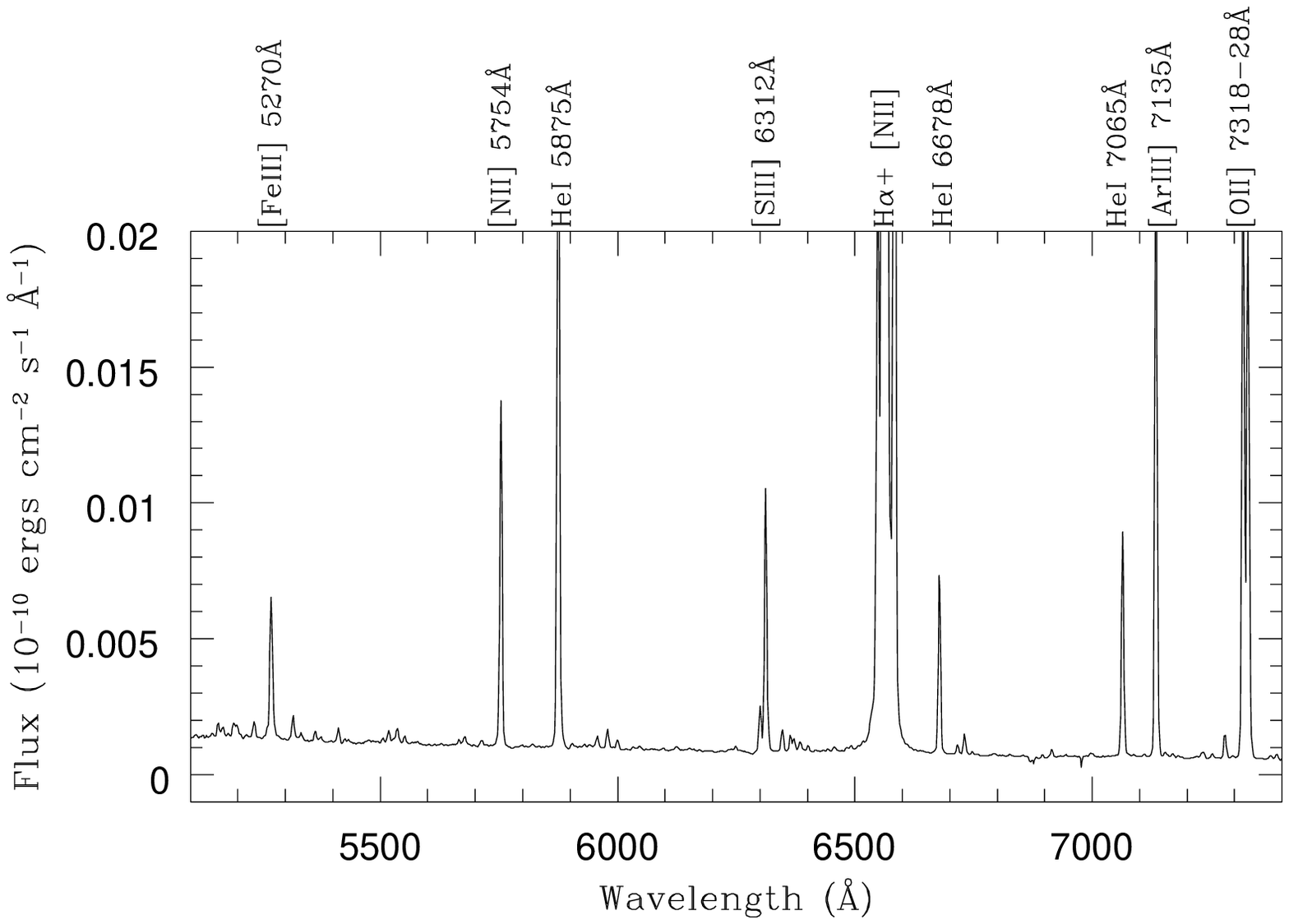}{9cm}{0}{70}{55}{-205}{0}
\caption{Parts of the low-resolution (B\&C) spectrum including the 
identification of the most prominent lines.}
\end{figure}

\section{Introduction}   

The classification of stars with respect to their evolutionary stage is not 
always straightforward. This is especially true for the southern galactic 
B[e] star Hen 2-90. It was first classified as a planetary nebula by  
\citet{Webster} and \citet{Henize}. Later, \citet{Costa} 
and \citet{Maciel} classified it as a planetary nebula with low metal 
abundance and with a central star of low mass and low luminosity.
\citet{Sahai00}, based on HST images, and 
\citet{Guerrero}, using ground-based images, have described the presence of a
nebula bisected by a disk and with both a bipolar jet and knots, spaced 
uniformly. The dynamical stability of the jets and knots makes Hen 2-90 even 
a unique object. These characteristics could point to a compact planetary 
nebula nature with the jets shaping the spherical AGB wind in a bipolar
scenario \citep{Sahai98,Imai,Vinkovic}
or to a binary nature like e.g. a symbiotic object \citep{Sahai00,Guerrero}.
Reason enough for us to take a closer look at this fascinating object by
studying its close-by circumstellar material by means of an analysis of its
forbidden emission lines arising in the optical spectrum.

\section{Observations}

High and low resolution observations were obtained with the Fiber-fed
Extended Range Optical Spectrograph (FEROS) and with the Boller \& Chivens
(B\&C) spectrograph, respectively, at the ESO 1.52m telescope in La Silla
(Chile, under agreement ESO/ON). Parts of the low resolution B\&C spectrum are shown in Fig.\,1,
and a line identification of the most prominent lines is given. 
Both, the fibre aperture of FEROS and the slit width of B\&C were 
2\arcsec~which is indicated by the circle overplotted on the HST image (top
panel of Fig.\,2). An extended description of the observations together with a 
classification of the line profiles and a complete line list is given in 
\citet{K2005}.

\begin{figure}[!ht]
\plotone{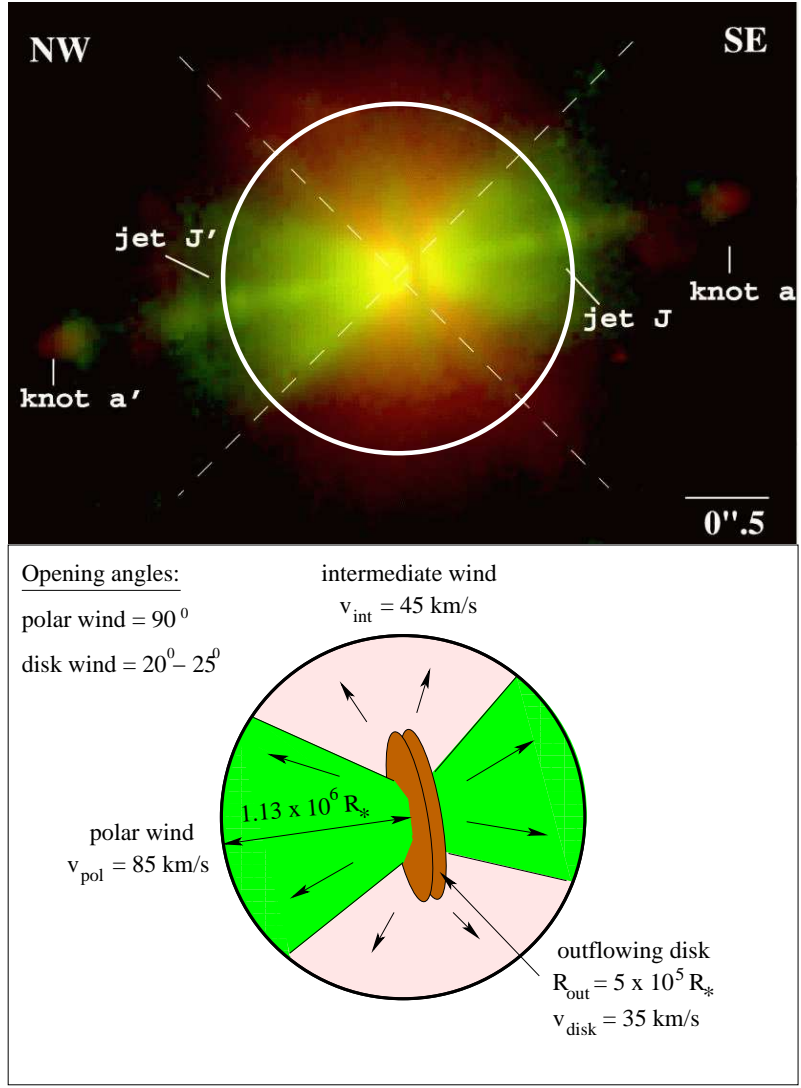}
\caption{Top: HST image of Hen 2-90 from \citet{Sahai02} indicating the 
bipolar wind separated by the equatorial edge-on seen disk-like structure. 
The big white circle indicates the aperture/slit width of our observations.
Bottom: Sketch of the non-spherical wind model divided up into the polar wind,
outflowing disk, and intermediate wind. Also given are the corresponding wind
terminal velocities, derived from the wings of the emission lines in the 
different wind parts.}
\end{figure}

\section{The nature of the circumstellar material of Hen 2-90}
                                             
There exist three major reasons why we think that the circumstellar material
can best be approximated by a non-spherical wind scenario:

{\bf 1. The emission lines:} Our spectra contain plenty of emission lines,
both permitted and forbidden. Usually, the density considerations for
these lines are completely different: while permitted lines need high density
regions to be strong, the forbidden lines need a low density environment.
The circumstellar medium of Hen 2-90 must therefore consist of at least two 
(probably more) different density regions.

{\bf 2. The ionization structure:} The HST image clearly shows dominance of
high-ionized material (O{\sc iii}) in the polar directions, lower ionized 
material at intermediate directions (N{\sc ii}), and a dark, probably neutral
disk-like structure in the equatorial plane. A zoo of different ionization 
states is also present in our optical spectra, and we want to stress that we 
even observed strong emission of [O{\sc i}], indicative for a huge amount of 
hydrogen neutral material in the vicinity of the central star and most probably 
associated with the dark disk-like structure.  

{\bf 3. The velocity structure:} From the high-resolution spectra we were
able to derive the wing velocities of the individual emission lines.
Interestingly, we found that the high-ionized lines have the highest velocities,
while the neutral lines (e.g. [O{\sc i}]) have the lowest velocities.

The combination of these three facts led us to the conclusion that the 
circumstellar medium of Hen 2-90 might best be modeled with a non-spherical 
wind scenario with latitude dependent temperature (ionization structure), 
density distribution, and outflow velocity. Such a scenario, as shown in the 
bottom panel of Fig.\,2, is the basis for our model calculations.

\section{The non-spherical wind model for Hen 2-90}

The stellar parameters of Hen 2-90 are not well known. For example, its
effective temperature ranges between 50\,000\,K \citep{Kaler,Costa} and 
32\,000\,K \citep{Cidale}, depending on the method used\footnote{Note, that 
classification as a B star is based on the emission spectrum, but it does not 
reflect the effective temperature of the star.}. During the modeling it turned 
out that the real surface temperature in the polar wind is not really
relevant, because the forbidden lines are produced further away from the star
where the wind temperature dropped already to its final value. 
Equally poorly known is the distance to Hen 2-90, which ranges between 1.5\,kpc 
\citep{Costa} and 2.5\,kpc \citep{Sahai00}, and we use a mean value of 2.0\,kpc.

\begin{table}[!ht]
\caption{Best fit model parameters for the polar, intermediate, and disk wind. 
The mass fluxes are assumed to be constant within the indicated $\theta$-ranges
and have an error on the order of 20\,\%.}
    \smallskip
      \begin{center}
        {\small
         \begin{tabular}{lrrrccc}
         \tableline
         \noalign{\smallskip}
Wind  & $T_{0}$  &  $T_{\infty}$ & $\theta_{\rm min}$ & $\theta_{\rm max}$ &
   $v_{\infty}$ & $F_{\rm m}(\theta)$ \\
 & [$10^{3}$\,K] & [$10^{3}$\,K] & [$\deg$] & [$\deg$] & [km\,s$^{-1}$] &
   [g\,s$^{-1}$cm$^{-2}$] \\
         \noalign{\smallskip}
         \tableline
         \noalign{\smallskip}
polar & 30--50 & 10 &  0 & 45 & 85 & $7.2\times 10^{-2}$ \\
intermediate & 30--50 & 10 & 45 & 78 & 45 & $1.5\times 10^{-1}$ \\
disk & 10 &  6 & 78 & 90 & 35 & $5.5\times 10^{-1}$ \\
         \noalign{\smallskip}
         \tableline
         \end{tabular}
        }
     \end{center}
  \end{table}

We concentrate our detailed analysis on the forbidden lines\footnote{For a
detailed description of the line luminosity calculation see \citet{K2005}.},
which are excellent indicators of the circumstellar material for two reasons:
                                                                                
\begin{itemize}
\item They are excited collisionally and are therefore sensitive to temperature
and density;
\item The circumstellar nebula is optically thin for these lines, simplifying
the analysis.
\end{itemize}
Available forbidden lines in our spectrum are from O{\sc i},
O{\sc ii}, O{\sc iii}, N{\sc ii}, Cl{\sc ii}, Cl{\sc iii}, Ar{\sc iii},
S{\sc ii}, and S{\sc iii}. We do not model lines from Fe{\sc ii} because
iron cannot be treated in such a simple way as the other elements.

The different ionization states mirror the different emission regions in the
non-spherical wind scenario and we could model the individual lines
self-con\-sis\-tently. The resulting parameters describing the non-spherical wind
are given in Table\,1. We find the following latitude dependences from pole
to equator:
\begin{itemize}
\item A decrease in surface temperature by a factor 3--5;
\item A decrease in terminal velocity by about a factor 2.5; 
\item An increase in mass flux by about a factor 8.
\end{itemize}
Such a variation of the different parameters with latitude strongly reminds
of the scenario of a rapidly rotating star. We used the decrease in terminal 
velocity to derive a rotation speed of $\sim 80$\% of the critical velocity 
(see Fig.\,6 of Kraus \& Lamers, this volume), which is on the same order as 
the values typically found e.g. for classical Be stars \citep{John}.

The total mass loss rate\footnote{Note, that this is the past mass loss rate 
during the superwind phase, not the present one.} was found to be on the order
of $3\times 10^{-5}\,M_{\odot}$yr$^{-1}$, which coincides nicely with those
found for proto-planetary nebulae with axially symmetric dust shells resulting
from a superwind phase \citep[see e.g.][]{Meixner}. 

Finally, we would like to mention that during the fitting of the observed
line luminosities we had to reduce the abundances of C, N, and O to 
values of $<0.6$, 0.5, and 0.3 solar, respectively, while Ar, S, and Cl
could be modeled with standard solar abundances. The depletions of C and O
follow from stellar evolution, but we have no explanation for the depletion
in N. There exist some B-type post AGB stars of similar behaviour in the halo 
of our galaxy \citep{Moehler}, and recently \citet{Lennon} reported on a Be
star with an unexpected low N abundance. The origin of those anomalies
is, however, still poorly understood.

\section{The nature of Hen 2-90}

In the literature, Hen 2-90 has been classified either as a (compact) planetary 
nebula \citep[e.g.][]{Costa,Maciel,Lamers}, or as a symbiotic object
\citep{Sahai00,Guerrero}. One goal of our study was also to clarify this
open question.
The major characteristics of a symbiotic object are:
\begin{itemize}
\item Emission of He{\sc ii} lines from the hot component;
\item Presence of TiO bands from the cool component. 
\end{itemize}
The absence of both features in our optical spectra clearly speaks against
the classification of Hen 2-90 as a symbiotic object. In addition, the lack
of any emission line from ions with ionization potential higher
than $\sim 40$\,eV even indicates that the effective temperature of Hen 2-90 
is more on the order of 30\,000\,K as suggested by \citet{Cidale}
than 50\,000\,K as stated by \citet{Kaler}.

On the other hand, the derived mass loss rate of $\sim 3\times 
10^{-5}\,M_{\odot}$yr$^{-1}$ and the strength of the [O{\sc iii}] 5007\AA~line 
(when taking into account the depletion in O) are consistent with what is
known for (compact) planetary nebulae, and even the jet-like structure and
regular blob ejection could be modeled perfectly with a planetary nebula
scenario and under the assumption of a solar-like magnetic cycle \citep{Garcia}.
 
From our observations, modeling and this discussion, we conclude that Hen 2-90
is at least an evolved object, and the classification as a compact planetary
nebula seems favourable. Whether it is indeed part of a binary could not
yet been proven \citep[see also][]{Krausetal05} and needs some further
investigation.

\section{Conclusions}

We studied the non-spherical mass loss history of the enigmatic B[e]
star Hen 2-90 by means of a detailed analysis of the optical forbidden 
emisison lines. The wind geometry used is based on the HST image, 
which reveals a non-spherical wind structure consisting of a high-ionized
polar wind, a neutral outflowing disk, and a low-ionized intermediate wind.
Such a latitude dependent ionization structure could be confirmed by  
the zoo of emission lines of different ions in our spectra. In addition, 
we found a decrease in velocity and an increase in mass flux from pole
to equator, which might be interpreted in terms of a rapidly rotating
(80\% critical) underlying star. Such rapid rotation might have caused
the massive non-spherical wind structure during a superwind phase with total
mass loss of about $3\times 10^{-5}\,M_{\odot}$yr$^{-1}$, as it is often 
observed during the formation of a (proto) planetary nebula. 
From our observations 
and modeling results we conclude that Hen 2-90 must be an evolved object,
and we favour the interpretation of a compact planetary nebula. Whether it 
is part of a binary system, is still an unsolved question. 

\acknowledgements 
M.K. acknowledges financial support from the Nederlandse Organisatie voor
      Wetenschappelijk Onderzoek grant No.\,614.000.310.
M.B.F. acknowledges financial support from \emph{CNPq} (Post-doc position -
150170/ 2004-1), \emph{Utrecht University}, \emph{LKBF} and \emph{NOVA}
foundations.



\vspace{-0.5cm}
\begin{thebibliography}{}

\bibitem[Cidale et al.(2001)]{Cidale} 
Cidale, L., Zorec, J., \& Tringaniello, L. 2001, A\&A, 368, 160
\bibitem[Costa et al.(1993)]{Costa} 
Costa, R.D.D., de Freitas Pacheco, J.A., \& Maciel, W.J. 1993, A\&A, 276, 184
\bibitem[Garc$\acute{\rm \i}$a-Segura et al.(2001)]{Garcia} 
Garc$\acute{\rm \i}$a-Segura, G., L\'opez, J.A., \& Franco, J.  2001, ApJ, 560, 928
\bibitem[Guerrero et al.(2001)]{Guerrero} 
Guerrero, M.A., Miranda, L.F., Chu, Y.H., Rodriguez, M., \& Williams, R.M. 2001, ApJ, 563, 883
\bibitem[Henize(1967)]{Henize} 
Henize, K.G. 1967, ApJS, 14, 125
\bibitem[Imai et al.(2002)]{Imai} 
Imai, H., Obara, K., Diamond, P.J., Omodaka, T., \& Sasao, T. 2002, Nature, 417, 829
\bibitem[Kaler \& Jacoby(1991)]{Kaler}
Kaler, J.B., \& Jacoby, G.H. 1991, ApJ, 372, 215
\bibitem[Kraus et al.(2005a)]{Krausetal05} 
Kraus, M., Borges Fernandes, M., \& de Ara\'ujo, F.X. 2005a, in Massive Stars in Interacting Binaries, ed. A. Moffat, \& N. St-Louis (San Francisco: ASP) {\sl astro-ph/0410196}
\bibitem[Kraus et al.(2005b)]{K2005}
Kraus, M., Borges Fernandes, M., de Ara\'{u}jo, F.X., \& Lamers, H.J.G.L.M. 2005b, A\&A, 441, 289
\bibitem[Lamers et al.(1998)]{Lamers} 
Lamers, H.J.G.L.M., Zickgraf, F.-J., de Winter, D., Houziaux, L., \& Zorec, J. 1998, A\&A, 340, 117
\bibitem[Lennon et al.(2005)]{Lennon} 
Lennon, D.J., Lee, J.-K., Dufton, P.L., \& Ryans, R.S.I. 2005, A\&A, 438, 265
\bibitem[Maciel(1993)]{Maciel} 
Maciel, W.J. 1993, Ap\&SS, 209, 65
\bibitem[Meixner et al.(1997)]{Meixner} 
Meixner, M., Skinner, C.J., Graham, J.R., Keto, E., Jernigan, J.G., \& Arens, J.F. 1997, ApJ, 482, 897
\bibitem[Moehler \& Heber(1998)]{Moehler} 
Moehler, S., \& Heber, U. 1998, A\&A 335, 985
\bibitem[Porter \& Rivinius(2003)]{John} 
Porter, J.M., \& Rivinius, Th. 2003, PASP, 115, 1153
\bibitem[Sahai \& Nyman(2000)]{Sahai00} 
Sahai, R., \& Nyman, L.-A. 2000, ApJ, 537, L\,145
\bibitem[Sahai \& Trauger(1998)]{Sahai98} Sahai, R., \& Trauger, J.T. 1998, AJ, 116, 1357
\bibitem[Sahai et al.(2002)]{Sahai02}
Sahai, R., Brillant, S., Livio, M., Grebel, E.K., Brandner, W., Tingay, S., \& Nyman, L.-A. 2002, ApJ, 573, L\,123
\bibitem[Vinkovi\'c et al.(2004)]{Vinkovic} 
Vinkovi\'c, D., Elitzur, M., Hofmann, K.-H., \&  Weigelt, G. 2004, in Asymmetric Planetary Nebulae III, Vol. XXX, ed. M. Meixner, J. Kastner, B. Balick, \& N. Soker (San Francisco: ASP)
\bibitem[Webster(1966)]{Webster} Webster, L.B. 1966, PASP, 78, 136
\end{thebibliography}
\end{document}